\input{aipcheck}

\documentclass[
    ,final            
  ]
  {aipproc}
\layoutstyle{6x9}

\begin{document}

\title{SDSS Observations of the Milky Way vs. N-body Models: 
 A Comparison of Stellar Distributions in the Position-Velocity-Metallicity Space}


\author{S. Loebman}{
   address={University of Washington, Dept. of Astronomy, Box 351580, Seattle, WA 98195}
}

\author{R. Ro\v{s}kar}{
   address={University of Washington, Dept. of Astronomy, Box 351580, Seattle, WA 98195}
}

\author{\v{Z}. Ivezi\'{c}}{
   address={University of Washington, Dept. of Astronomy, Box 351580, Seattle, WA 98195}
}

\author{M. Juri\'{c}}{
  address={Institute for Advanced Study, 1 Einstein Drive, Princeton, NJ 08540}
}

\author{B. Sesar}{
 address={University of Washington, Dept. of Astronomy, Box 351580, Seattle, WA 98195}
}

\author{V. P. Debattista}{
 address={RCUK Fellow at Centre for Astrophysics, University of Central Lancashire, 
 	Preston, PR1 2HE, UK}
}
\author{T. R. Quinn}{
 address={University of Washington, Dept. of Astronomy, Box 351580, Seattle, WA 98195}
}

\author{G. S. Stinson}{
 address={Department of Physics and Astronomy, McMaster University, 
	Hamilton, ON, L8S 4M1, Canada}
}

\author{J. Wadsley}{
address={Department of Physics and Astronomy, McMaster University, 
	Hamilton, ON, L8S 4M1, Canada}
}

\begin{abstract}
The data obtained by the recent modern sky surveys enable detailed studies of
the stellar distribution in the multi-dimensional space spanned by spatial 
coordinates, velocity and metallicity, from the solar neighborhood all the
way out to the outer Milky Way halo. While these results represent exciting
observational breakthroughs, their interpretation is not simple. For example, 
traditional decomposition of the thin and thick disks predicts a strong
correlation in metallicity and kinematics at $\sim$1 kpc from the Galactic
plane; however, recent SDSS--based work has demonstrated an absence of this
correlation for disk stars. Instead, the variation of the metallicity and
rotational velocity distributions can be modeled using non--Gaussian functions 
that retain their shapes and only shift as the distance from the mid--plane 
increases. To fully contextualize these recent observational results, a
detailed comparison with sophisticated numerical models is necessary. Modern 
simulations have sufficient resolution and physical detail to study the
formation of stellar disks and spheroids over a large baseline of masses and 
cosmic ages. We discuss preliminary comparisons of various observed maps and 
N--body model predictions and find them encouraging. In particular, the 
N--body disk models of Ro\v{s}kar et al. \cite{Roskar 2008} reproduce a change of
disk scale height reminiscent of thin/thick disk decomposition, as well
as metallicity and rotational velocity gradients, while not inducing a
correlation of the latter two quantities, in qualitative agreement with 
SDSS observations.
\end{abstract}
\maketitle


\section{ Introduction }

Most studies of the Milky Way structure can be described as investigations 
of the stellar distribution in the nine--dimensional space spanned by the 
three spatial coordinates, three velocity components, and three main stellar 
parameters (luminosity, effective temperature, and metallicity). Recently,
several SDSS--based studies have provided unprecedented observational 
 constraints in various projections of this nine--dimensional space.
Juri\'{c} et al. (\cite{Juric 2008}, hereafter J08) used photometric parallax method 
to estimate distances to $\sim$48 million stars, and studied their spatial 
distribution. Thanks to accurate SDSS photometry that enabled reasonably accurate 
distances (10--15\%, \cite{Sesar 2008}), faint magnitude limits ($r<22$), and a 
large sky coverage (6500 deg$^2$), J08 were able to robustly constrain 
the parameters of a model for global spatial distribution of stars in the 
Milky Way. While their model is qualitatively similar to previous work 
exemplified by the Bahcall \& Soneira \cite{Bahcall 1980} model, J08 detected abundant 
substructure, and a clear change of slope in the counts of disk stars 
as a function of distance from the Galactic plane, usually interpreted 
as the transition from the thin to thick disk \cite{Gilmore 1983}. 
Ivezi\'{c} et al. (\cite{Ivezic 2008}, hereafter I08) further extended global analysis 
of SDSS data by developing a photometric metallicity estimator, and by 
utilizing a large proper motion catalog based on SDSS and Palomar
Observatory Sky Survey data \cite{Munn 2004}. I08 studied the dependence
of metallicity and rotational velocity for disk stars on the distance 
from the Galactic plane and detected gradients of both quantities over 
the distance range from several hundred pc to several kpc.
Such gradients would be expected in the traditional thin/thick disk
decomposition where the thick disk stars have a well defined bulk rotational
velocity lag and lower metallicity compared to those of the thin disk.
However, such a model would also also predict a correlation between
metallicity and the velocity lag, which is strongly excluded ($\sim7\sigma$
level) by the I08 analysis.
More sophisticated models are therefore needed. These, at minimum, have to
answer the following questions:
\begin{enumerate} 
\item Do the models reproduce the change of slope in the counts of disk 
stars as a function of distance from the Galactic plane?
\item If so, do they reproduce the gradients in metallicity and
     rotational velocity? 
\item If so, are metallicity and rotational velocity uncorrelated? 
\end{enumerate}
If the answer to all these questions is yes, then the model may be used to 
extract further insight about the importance of various physical mechanisms
operating in the disk of the Milky Way, and in particular it might allow us to study 
correlations of observables with stellar age and other quantities that 
are hard or impossible to measure directly.

\section{Simulation}

Here we analyze the results of an $N$--body + Smooth Particle Hydrodynamics (SPH) 
simulation designed to mimic the quiescent formation and evolution of a Milky Way--type galactic disk  
following the last major merger. The system is initialized as in Kaufmann et al. \cite{Kaufmann 2007} and 
Ro\v{s}kar et al. (\cite{Roskar 2008},\cite{Roskar 2008a}, hereafter R08 and R08a respectively), and consists of a 
rotating, pressure--supported 
gas halo embedded in an NFW \cite{Navarro 1997} dark matter halo. As the simulation proceeds,
the gas cools and collapses to the center of the halo, forming a thin disk from the inside--out. 
When the gas reaches densities and temperatures conducive to star formation, the sub--grid 
star formation and stellar feedback recipes are initiated \cite{Stinson 2006}. Importantly, 
the stellar feedback prescriptions include SN II, SN Ia and AGB metal production, as well 
as injection of supernova energy which impacts the hydrodynamic properties of the disk ISM. Note that 
we make no {\it a priori} assumptions about the disk's structure -- its growth and the subsequent
evolution of its stellar populations are completely spontaneous and governed only by 
hydrodynamics,stellar feedback, and gravity. Although we do not account for the full cosmological context, merging
in the $\Lambda$CDM paradigm is a higher order effect at the epochs in question \cite{Brook 2005}. By 
simplifying our assumptions, we are able to use higher resolution and more easily study 
the impact of key dynamical effects to observational properties of stellar populations.

Based on these simulations, R08 and R08a presented the implications of 
stellar radial migration resulting from the interactions of stars with transient spiral arms 
\cite{Sellwood 2002} on the observable properties of disk stellar populations. 
Here we  explore whether the picture of disk evolution presented in R08 and R08a may 
also help understand the correlations (and lack thereof) in the disk properties derived 
from SDSS data. 


\section{ Analysis and  Discussion  } 
				   
To be consistent with the analysis of high galactic latitude SDSS data
by J08 and I08, we select model particles from an annulus with 7 kpc $<R<$ 9 kpc, where
$R$ is the galactocentric cylindrical radius. We study the stellar mass distribution, rotational
velocity and metallicity as functions of distance from the Galactic plane, $Z$.
The behavior of model results in illustrated in Figure 1. 

We find that the model distribution of stars as a function of $Z$ resembles
a sum of two exponential profiles, with the ``break'' height of $Z\sim1.1$ kpc. 
The best-fit scale heights are 365 pc and 500 pc, in qualitative agreement with SDSS
data \cite{Juric 2008}. Quantitatively, the scale height ratio suggested by data is
higher ($\sim$3, instead of $\sim$1.4) and the ``thick disk'' normalization 
is $\sim$0.13, rather than $\sim0.3$. Nevertheless, since these models are 
not specifically tuned to reproduce the Milky Way, we find this agreement 
remarkable. 

The age distribution as a function of $Z$ shows that only very old stars are found 
at large $Z$: the median age is $\sim$7 Gyr at $Z\sim1$ kpc. The rotational velocity 
depends on age: the older a star is, the slower is its rotation. Together with the 
change of age distribution with $Z$, this correlation leads to a $Z$ gradient of 
rotational velocity: the best-fit value is 15 km/s, in qualitative agreement
with the measured value of $\sim$30 km/s \cite{Ivezic 2008}. 

The metallicity distribution also changes with $Z$: the best-fit gradient
is $\sim$0.12 dex/kpc, again in qualitative agreement with the measured value 
of $\sim$0.3 dex/kpc \cite{Ivezic 2008}. Although both rotational velocity and metallicity 
show vertical gradients, when stars are selected from a thin $Z$ slice, 
velocity and metallicity are not correlated. Hence, the models of R08 
are in qualitative agreement with the data and provide affirmative answers
to all three questions posed in Introduction, at least in a qualitative 
sense. 

In summary, the models of R08 show remarkable similarity to SDSS observations 
of the distribution of Milky Way stars in the position--velocity--metallicity 
space. In addition, these models provide a quantity that SDSS observations 
cannot -- the stellar age.  We intend to further explore the distribution of 
stellar age and its correlations to various observables in future work.

\begin{figure}[!H]
\includegraphics[height=.3\textwidth,width=0.45\textwidth]{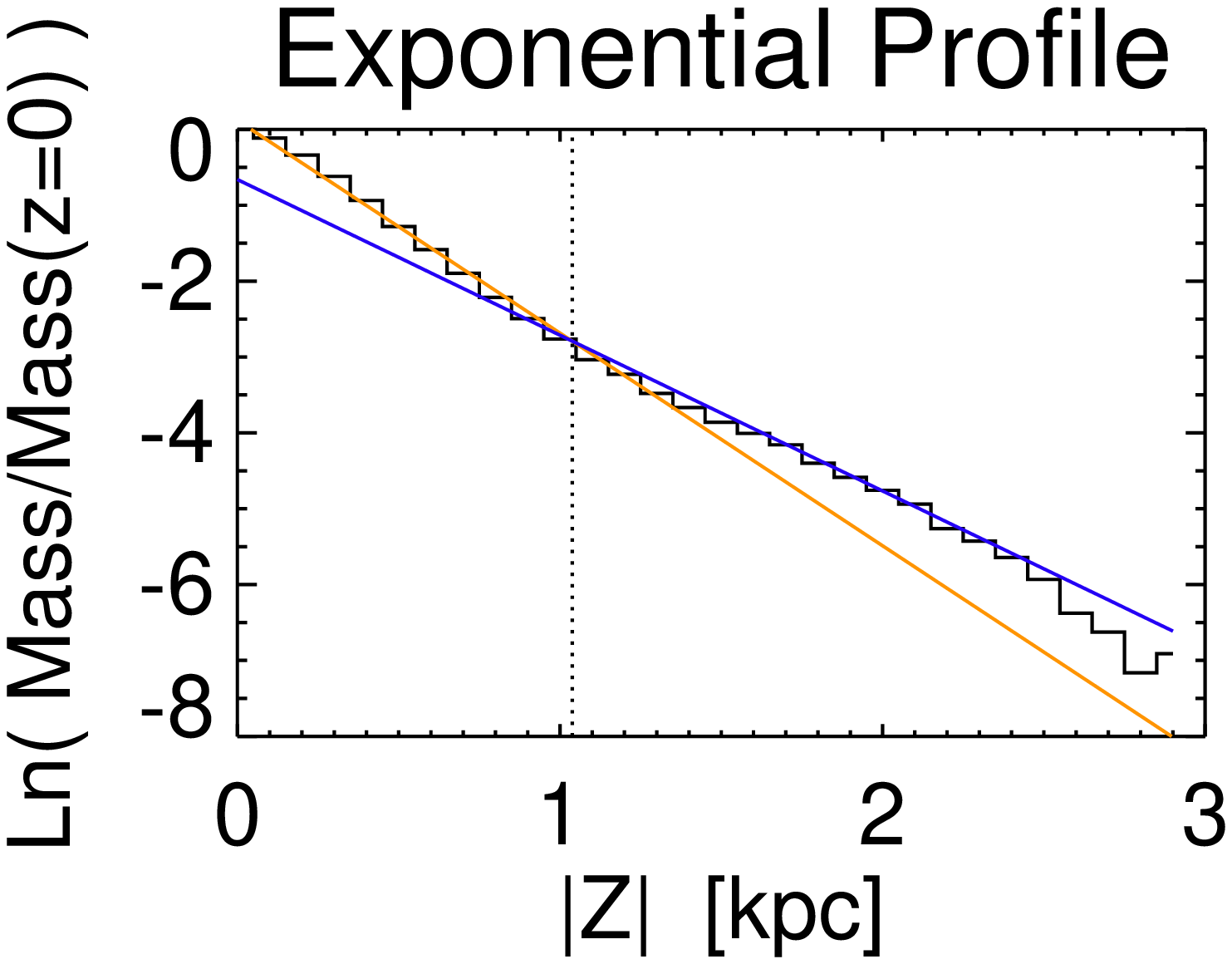}
\includegraphics[height=.3\textwidth,width=0.45\textwidth]{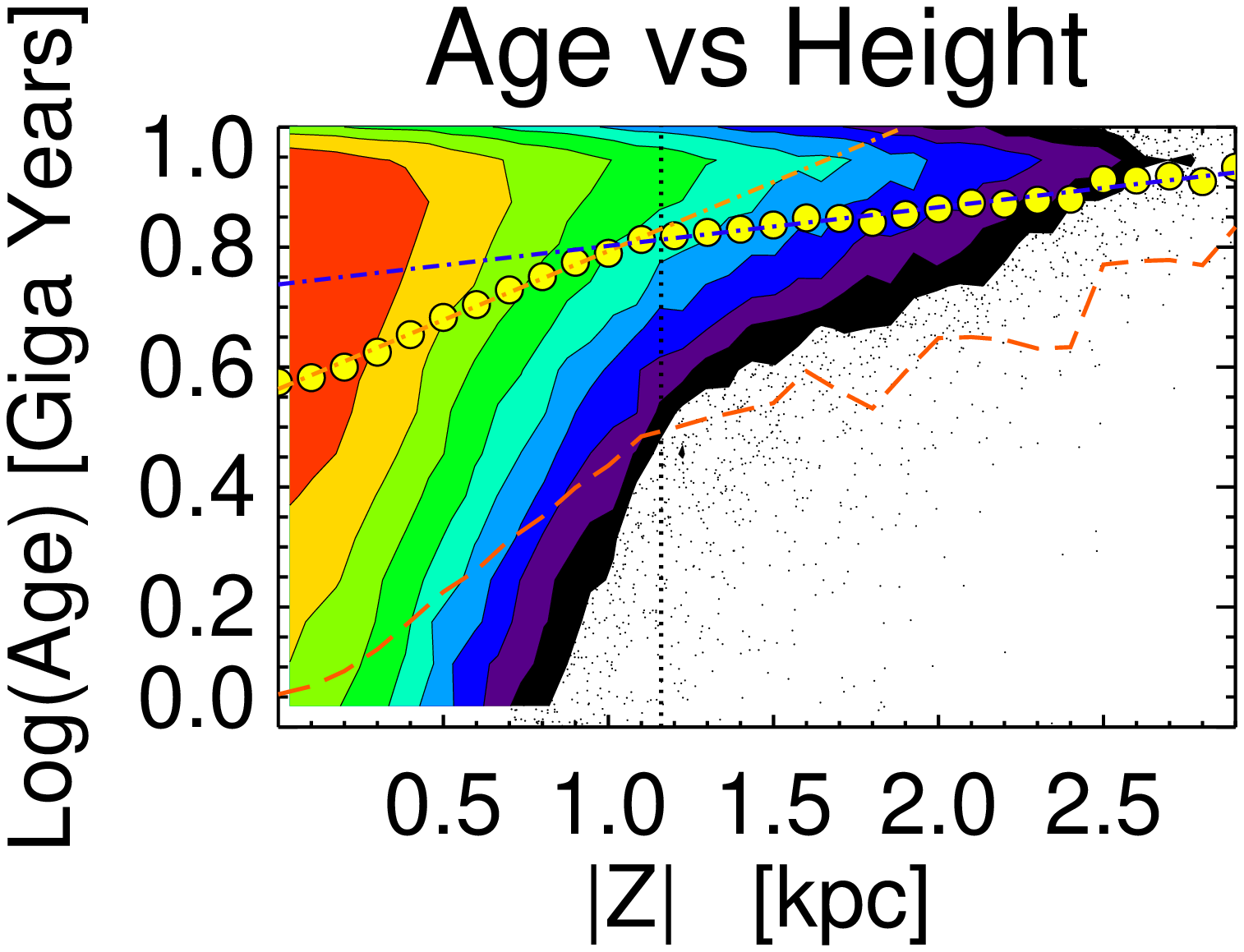}
\end{figure}

\begin{figure}[!h!t]
\includegraphics[height=.3\textwidth,width=0.45\textwidth]{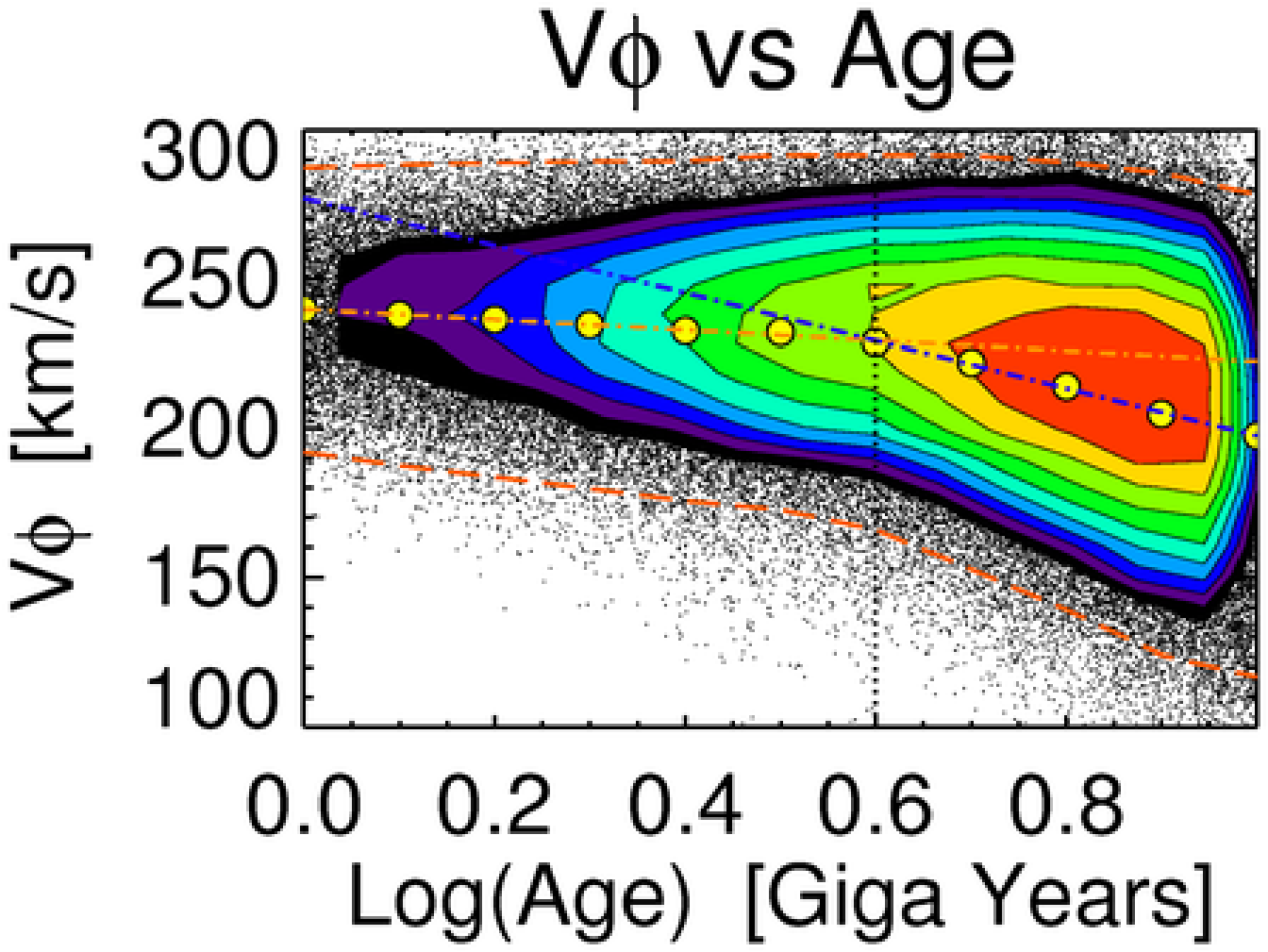}
\includegraphics[height=.3\textwidth,width=0.45\textwidth]{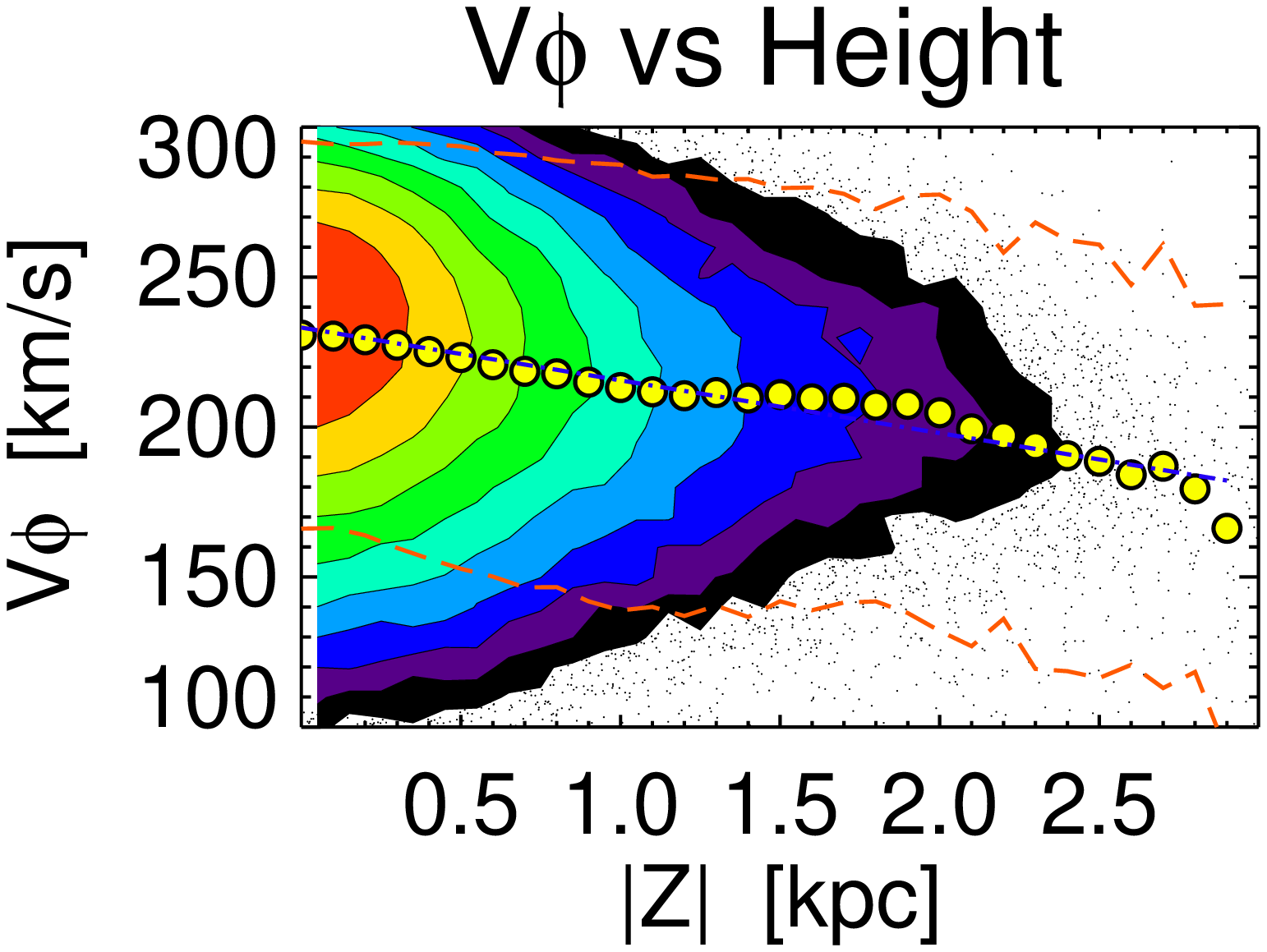}
\end{figure}

\begin{figure}[!h!t]
\includegraphics[height=.3\textwidth,width=0.45\textwidth]{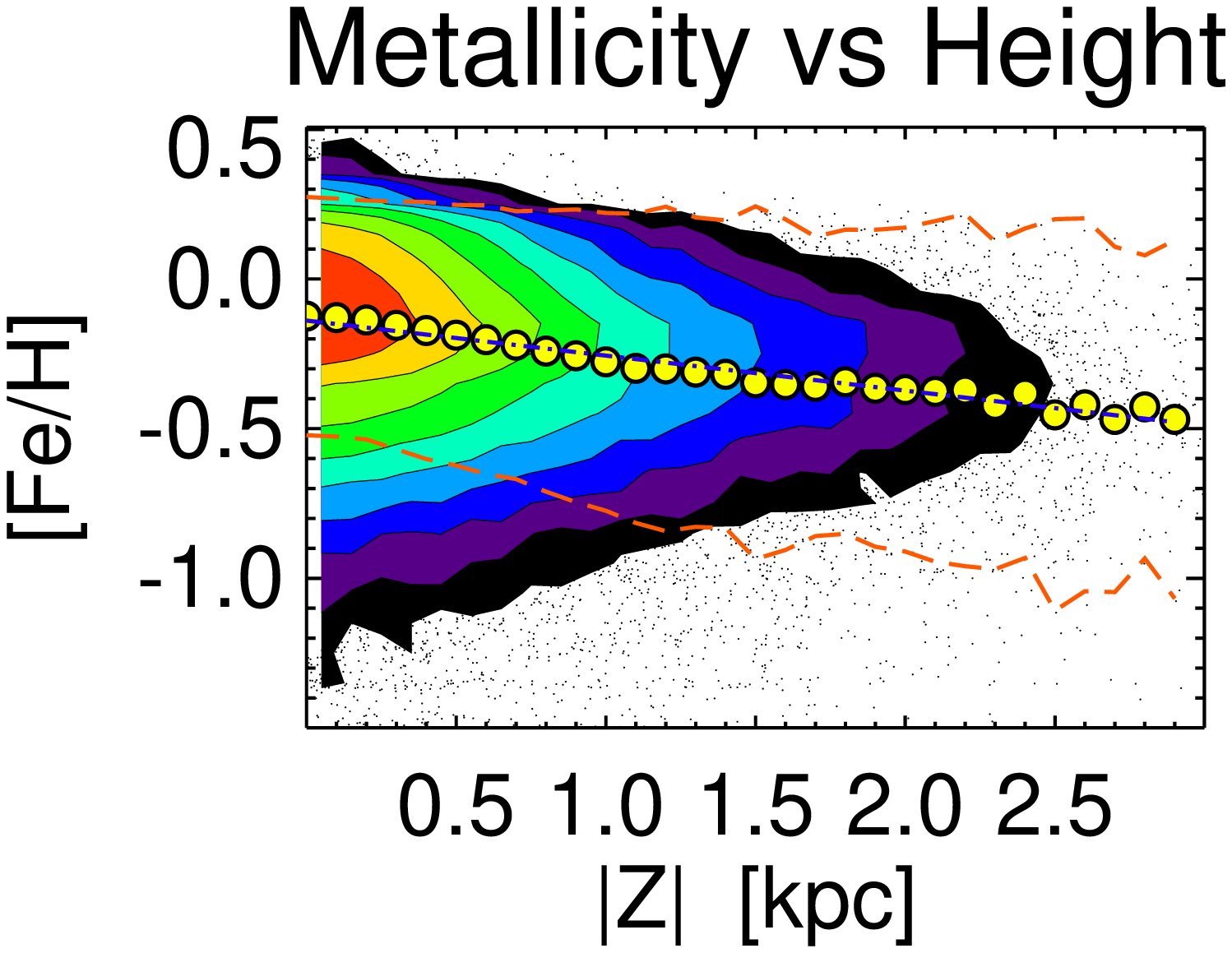}
\includegraphics[height=.3\textwidth,width=0.45\textwidth]{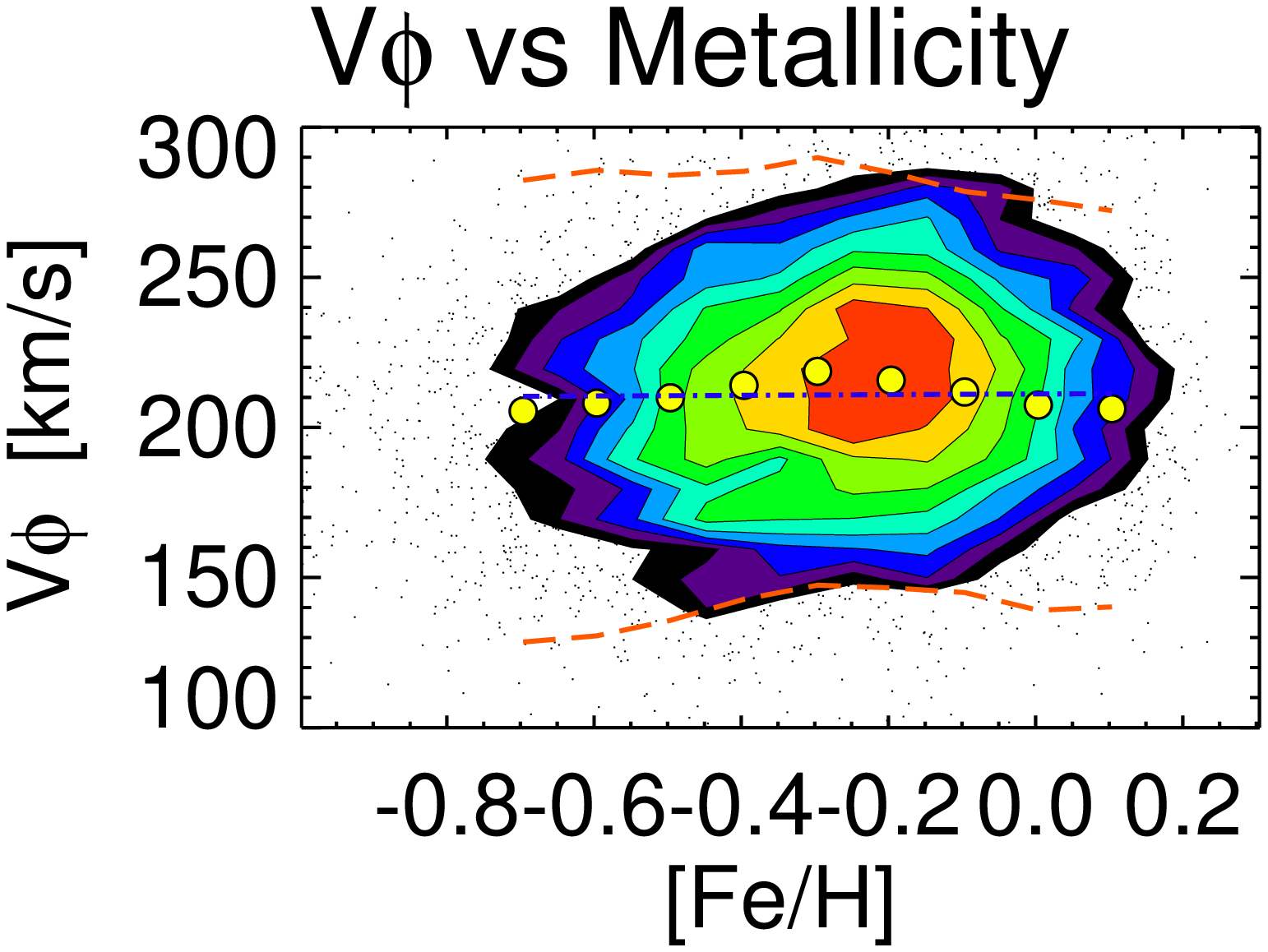}
\caption{
The behavior of $\sim$200,000 model particles selected from a galactocentric cylindrical
annulus with 7 kpc $<R<$ 9 kpc. The histogram in the top left panel shows
mass-weighted counts as a function of distance from the Galactic plane, $Z$.
The two solid straight lines show the best-fit thin and thick exponential density 
profiles, and the vertical dashed line marks the transition. The age distribution 
as a function of $Z$ is shown in the top right panel as color-coded contours 
(low to medium to high: black to green to red) in the regions of high density
of points, and as individual points otherwise. The large symbols show the median
values in $Z$ bins, and the dashed lines show a 2$\sigma$ envelope around the
medians. The dot-dashed line shows the best linear fit to these medians. 
The remaining four panels are analogous, except that they show the
rotational velocity vs. age (middle left),  the rotational velocity vs. $Z$ 
(middle right), the metallicity vs. $Z$ (bottom left) and the rotational 
velocity vs. metallicity (bottom right) diagrams. In the last panel, 
only a subset of data from a thin slice in $Z$ centered at $\sim$1 kpc 
is shown. Note the absence of a correlation between the velocity
and metallicity. 
}
\end{figure}


\begin{theacknowledgments}
We thank our numerous SDSS (www.sdss.org) collaborators for their 
valuable contributions and helpful discussions.
 
This research was supported in part by the NSF through TeraGrid resources provided by TACC and PSC.  R.R. and T.R.Q. were supported by the NSF ITR grant PHY-0205413 at the University of Washington.
\end{theacknowledgments}

\bibliographystyle{aipproc}   


\IfFileExists{\jobname.bbl}{}
 {\typeout{}
  \typeout{******************************************}
  \typeout{** Please run "bibtex \jobname" to optain}
  \typeout{** the bibliography and then re--run LaTeX}
  \typeout{** twice to fix the references!}
  \typeout{******************************************}
  \typeout{}
 }


\end{document}